\documentclass[english,preprint]{revtex4}
\usepackage[T1]{fontenc}
\usepackage[latin9]{inputenc}
\usepackage{graphicx}

\usepackage{babel}

\begin{document}

\title{Delay and periodicity}

\author{S. Yanchuk$^{1}$ and P. Perlikowski$^{1,2}$}

\affiliation{$^{1}$Humboldt University of Berlin, Institute of Mathematics, Unter
den Linden 6, 10099 Berlin, Germany; \\
$^{2}$Division of Dynamics, Technical University of Lodz, Stefanowskiego
1/15, 90-924 Lodz, Poland}

\begin{abstract}
Systems with time delay play an important role in modeling of many
physical and biological processes. In this paper we describe generic
properties of systems with time delay, which are related to the appearance
and stability of periodic solutions. In particular, we show that delay
systems generically have families of periodic solutions, which are
reappearing for infinitely many delay times. As delay increases, the
solution families overlap leading to increasing coexistence of multiple
stable as well as unstable solutions. We also consider stability issue
of periodic solutions with large delay by explaining asymptotic properties
of the spectrum of characteristic multipliers. We show that the spectrum
of multipliers can be splitted into two parts: pseudo-continuous and
strongly unstable. The pseudo-continuous part of the spectrum mediates
destabilization of periodic solutions.
\end{abstract}
\maketitle

\section{Introduction}

The dynamical behavior of various physical and biological systems
under the influence of delayed feedback or coupling can be modeled
by including terms with delayed arguments in the equations of motion.
When the delay becomes longer than the other characteristic timescales
of the system, a complicated and high-dimensional dynamics can appear
\citep{Giacomelli1996,Mensour1998,Wolfrum2006,Bestehorn2000,Fischer2000,Giacomelli1998}.
The analysis and control of such dynamical regimes is important for
many applications including lasers with optical feedback or coupling
\citep{Fischer2000,Goedgebuer1998,Fiedler2008}, neural activity control
\citep{Tass2008,Popovych2005}, and many others \citep{Scholl2008}.
For instance, the following complicated regimes have been observed
in lasers with delayed feedback: low-frequency fluctuations \citep{Heil1998},
regular pulse packages \citep{Gavrielides1999}, coherence collapse,
just to mention a few. 

One of the fundamental question in the analysis of systems with delay
concerns properties of periodic solutions. Periodic solutions of any
dynamical system, including also systems with delay, are important
parts of the dynamics. When such solutions are stable, they can be
directly observed experimentally or numerically. In the case, when
such solutions are unstable, they play an important role, e.g. by
determining of a set of admissible initial values to be attracted
to some stable steady state (basin boundary), or by forming fundamental
blocks of a chaotic attractor \citep{Auerbach1987,Yanchuk2001}. Finally,
unstable as well as stable periodic solutions can play an important
role in mediating any kind of soft or hard transitions as some control
parameters are varied. 

This paper is devoted to the study of generic properties of periodic
solutions in systems with a constant time delay \begin{equation}
x'(t)=f(x(t),x(t-\tau)),\label{eq:dde}\end{equation}
where $x\in R^{n}$ and $\tau>0$ is the time delay. Since we investigate
the influence of the delay, we assume $\tau$ to be our control parameter.
Delay has been used as a parameter in various applications: chaotic
systems with feedback \citep{Ikeda1982,Ikeda1987}, network motifs
\citep{Schuster1989,DHuys2008}, large networks or arrays of oscillators
with delayed coupling \citep{Kim1997,Montbrio2006,Dodla2004,Kinzel},
mechanical systems \citep{Campbell1995,Xu2003,Yamasue2004,Stefanski2005},
laser systems with feedback \citep{Simonet1995} and coupling \citep{Vicente2006},
coupled neurons \citep{Rossoni2005,Masoller2008}, chemical oscillators
\citep{Erneux2008}, delayed feedback control \citep{Fiedler2007,Fiedler2008,Just2007,Balanov2005,Feito2007,Ahlborn2005,Ryu2004,Popovych2005,Hovel2005,Postlethwaite2007}.
We believe that our results are applicable to the all above mentioned
cases as well as to many others, which include time delay as a controllable
parameter. 

The plan of the paper is as follows. Section \ref{sec:Reappearance-of-periodic}
starts by showing that periodic solutions in system (\ref{eq:dde})
are forming branches with respect to the control parameter $\tau$.
These branches are reappearing infinitely many times for different
delay values. As the delay increases, the solution branches overlap
leading to increasing coexistence of multiple stable as well as unstable
periodic solutions. The number of coexisting solutions is shown to
be linearly proportional to the delay time. Further in Section \ref{sec:Stability-of-the},
we consider stability properties of periodic solutions for systems
with large delay by explaining asymptotic properties of the spectrum
of their characteristic multipliers. The spectrum of multipliers can
be splitted into two parts: pseudo-continuous and strongly unstable.
Such situation is similarl to the case of steady states in delay systems
with large delay \citep{Wolfrum2006,Yanchuk2006a,Yanchuk2005a}. The
pseudo-continuous spectrum mediates bifurcations of periodic solutions
for systems with large delay. Sections \ref{sec:Asymptotic-stability-of}
and \ref{sec:Computation-of-pseudo-continuous} discuss some implications
of the existence of pseudo-continuous spectrum and possibility for
its numerical computation.

The obtained results provide a better understanding of mechanisms
behind the growing multistability and dynamic complexity in systems
with delay. In particular, we show that coexistence of multiple stable
(as well as unstable) periodic solutions is a natural feature of delay
systems. The growing {}``effective dimension'' of dynamics with
the growing delay is supported by the fact that the dimensionality
of unstable manifolds of periodic solutions also grows linearly with
delay. Our results will be illustrated using Duffing oscillator with
delay \begin{equation}
x''(t)+dx'(t)+ax(t)+x^{3}(t)+b\left(x(t)-x(t-\tau)\right)=0,\label{eq:duffing}\end{equation}
where $x(t)$ is a real variable, and Stuart-Landau oscillator with
delayed feedback \begin{equation}
z'(t)=(\alpha+i\beta)z(t)-z(t)|z(t)|^{2}+\kappa z(t-\tau),\label{eq:SL0}\end{equation}
where $z(t)$ is a complex variable.

\section{Reappearance of periodic solutions and branches \label{sec:Reappearance-of-periodic}}

In this section we show that periodic solutions of system (\ref{eq:dde})
reappear infinitely many times for different and well specified values
of delay $\tau$. We will also show that such solutions typically
form branches, which can be mapped one onto another by some similarity
transformation.

\subsection{Reappearance of periodic solutions}

\begin{figure}
\includegraphics{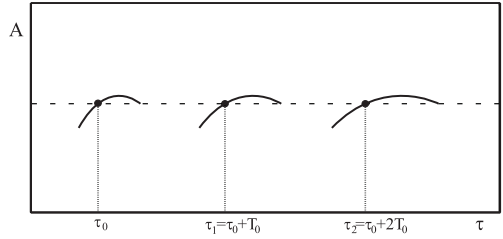}

\caption{\label{fig:Mapping-of-periodic}Reappearance of a periodic solution
$x_{0}(t)$ for the delays $\tau_{n}=\tau_{0}+nT_{0}$, $n=1,2,3,\dots$.
$T_{0}$ is the period of $x_{0}(t)$.}

\end{figure}

Let us consider system (\ref{eq:dde}), which possesses a periodic
solution $x_{0}(t)$ for a time delay $\tau=\tau_{0}$. Let $T_{0}$
be the period of this solution. Then it is easy to check that the
same periodic solution exists in system (\ref{eq:dde}) with time
delay $\tau_{1}=\tau_{0}+T_{0}$. Indeed, substituting $x_{0}(t)$
into (\ref{eq:dde}) we obtain \[
x_{0}'(t)=f(x_{0}(t),x_{0}(t-\tau_{1}))=f(x_{0}(t),x_{0}(t-\tau_{0}-T_{0}))\]
\[
=f(x_{0}(t),x_{0}(t-\tau_{0})),\]
where the periodicity of $x_{0}(t)$ is taken into account. Similarly,
the solution $x_{0}(t)$ reappears for all values $\tau_{n}=\tau_{0}+nT_{0}$,
$n=1,2,3,\dots$ of the delay; see Fig.~\ref{fig:Mapping-of-periodic}. 

\subsection{Reappearance of branches of periodic solutions}

A periodic solution $x_{0}(t)$ can be generically continued to a
branch of periodic solutions $x_{0}(t;\tau)$ with respect to the
parameter $\tau$, at least in some neighborhood of $\tau_{0}$. Denote
$T(\tau)$ to be the period of these solutions along the branch. Since
each individual periodic solution reappears for every delay time $\tau+nT(\tau)$,
the whole branch will appear infinitely many times as well; see Fig.~\ref{fig:Mapping-of-periodic}.
It is naturally to introduce the notion of \emph{primary branch},
which satisfies $\tau<T(\tau)$. For convenience, let us parametrize
the primary branch using the parameter $l$, which coincides with
the delay on this branch $x_{0}(t;l):=x_{0}(t;\tau)$. A solution
$x_{0}(t;l)$, which corresponds to some parameter value $l$, will
appear again on the $n$-th branch at time delay $\tau(n,l)=l+nT(l)$;
see Fig.~\ref{fig:Mapping-of-periodic}. Thus, we obtain the representation
of the $n$-th branch \begin{equation}
x_{n}(t;\tau(n,l))=x_{n}(t;l+nT(l))=x_{0}(t;l).\label{eq:nvia0branch}\end{equation}
The corresponding mapping, which maps delay times, is given as follows
\begin{equation}
l\to\tau(n,l)=l+nT(l).\label{eq:Tmap}\end{equation}

Examples of the above described branches can be numerously found in
the research literature \citep{Erneux2008,Dodla2004,Xu2003,Kim1997,Feito2007,DHuys2008,Ikeda1980,Ikeda1982}.
Usually, these branches can be found numerically. A useful tools for
finding such branches is the continuation software DDE-Biftool \citep{Engelborghs2001}.
In fact, only the primary branch should be computed while the others
can be obtained using the transformation (\ref{eq:nvia0branch}).

\begin{figure}
\includegraphics[width=8cm]{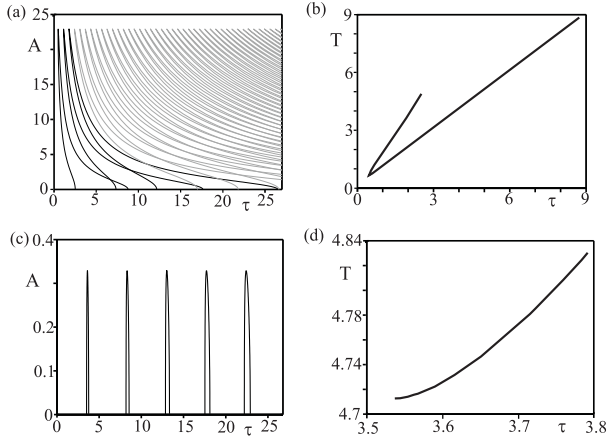}

\caption{Branches of periodic solutions for Duffing oscillator. \label{fig:Branches-of-periodic}
Left panel shows the amplitude of the solutions versus time delay,
right panel shows dependence of the period along the branch. Parameter
values: (a,b) $a=0.5$,\textbf{ $b=0.6$, $d=0.06$;} (c,d) $a=1.38$,\textbf{
$b=0.4$, $d=0.3$.} The first three branches are plotted in black and the rest in gray.}

\end{figure}

For the model of Duffing oscillator (\ref{eq:duffing}), we have found
two types of branches, which are presented in Fig.~\ref{fig:Branches-of-periodic}.
Other examples are given in Fig.~\ref{fig:SL}. Note that due to
the nontrivial dependence of the period $T(\tau)$ along the branch
(see the right panel of the figure), the mapping (\ref{eq:Tmap})
is not just a parallel shift. It has some more complicated properties,
which will be studied in the following subsection.

\subsection{Properties of the branches}

As we have seen, the primary branch of periodic solutions $x_{0}(t;l)$
is characterized by a period function along the branch $T(l)$. Clearly,
since all other branches consist of the same solutions, they have
the same period dependence $T(t)$. Via the mapping (\ref{eq:Tmap}),
the function $T(l)$ determines uniquely how branches reappear for
larger delay times. Let us discuss main properties of the map (\ref{eq:Tmap})
and corresponding implications.

\subsubsection*{Stretching, squeezing. }

Under the transformation (\ref{eq:Tmap}) some parts of the branch
will be stretched and some parts squeezed. In particular if \begin{equation}
\left|\frac{\partial\tau(n,l)}{\partial l}\right|=|1+nT'(l)|>1\label{eq:streching}\end{equation}
then the corresponding part will be locally stretched. Here $T'(l)=dT(l)/dl$
is the derivative of the period function. If the inverse inequality
\begin{equation}
\left|\frac{\partial\tau(n,l)}{\partial l}\right|=|1+nT'(l)|<1\label{eq:squeezing}\end{equation}
is satisfied, the corresponding branch parts will be locally squeezed.
With the increasing of $n$ (which is equivalent to increasing $\tau$)
almost all parts of the branches will be stretched, since (\ref{eq:streching})
will be satisfied. Hence, the branches become eventually wider and
occupy larger $\tau$ intervals. This leads to the growing overlapping
of branches and growing co-existence of periodic solutions with increasing
$\tau$. This effect is clearly visible in Figs.~\ref{fig:Branches-of-periodic}(a)
and \ref{fig:SL}.

\subsubsection*{\textmd{Multistability.} }

Let us describe the above mentioned phenomenon on a more quantitative
level. We will show that the number of coexistent periodic solutions
grows linearly with $\tau$ and give estimations for the corresponding
coefficient. Let us consider the two possible cases:

CASE 1. The primary branch is confined to some interval of $\tau$
as in the case shown in Fig.~\ref{fig:Branches-of-periodic}(a-d).
This means that the primary branch ranges from $l_{\min}$ till $l_{\max}$
($l_{\min}$ can be zero).

CASE 2. The primary branch is bounded only from below, i.e. it ranges
from $l_{\min}$ till $+\infty$; see an example in Fig.~\ref{fig:SL}(a).

Consider the first case. Let us denote $T_{\max}$ and $T_{\min}$
to be the maximum and the minimum of the period function $T(l)$ on
the interval $l_{\min}\le l\le l_{\max}$. If $T_{\max}=\infty$ then
the problem can be reduced to the Case 2, since the next branch $x_{1}(t;l+T(l))=x_{0}(t;l)$
will be stretched up to $\tau=\infty$ by the mapping (\ref{eq:Tmap}).
Hence, $T_{\max}$ and $T_{\min}$ can be considered to be bounded.
In this case, all other branches are also bounded and exist for delay
values $\tau(n,l)=l+nT(l)$, $l_{\min}\le l\le l_{\min}$. In particular,
the $n$-th branch ranges from\[
\tau_{\min}(n,l)=\min_{l_{\min}\le l\le l_{\max}}\tau(n,l)=\min_{l_{\min}\le l\le l_{max}}\left[l+nT(l)\right]\]
 till \[
\tau_{\max}(n,l)=\max_{l_{\min}\le l\le l_{\max}}\tau(n,l)=\max_{l_{\min}\le l\le l_{max}}\left[l+nT(l)\right].\]
For large enough $n$, the minimal and maximal bounds of the branches
can be well approximated as follows\begin{equation}
\tau_{\min}(n,l)=n\min\left[\frac{l}{n}+T(l)\right]\approx nT_{\min},\label{eq:Taumin}\end{equation}
\begin{equation}
\tau_{\max}(n,l)=n\max\left[\frac{l}{n}+T(l)\right]\approx nT_{\max}\label{eq:Taumax}\end{equation}
up to the terms of order one. Let us now find how many branches are
overlapping for some sufficiently large delay value $\tau$. It is
clear that these branches should satisfy the condition \[
\tau_{\min}(n,l)<\tau<\tau_{\max}(n,l).\]

\begin{figure}
\includegraphics{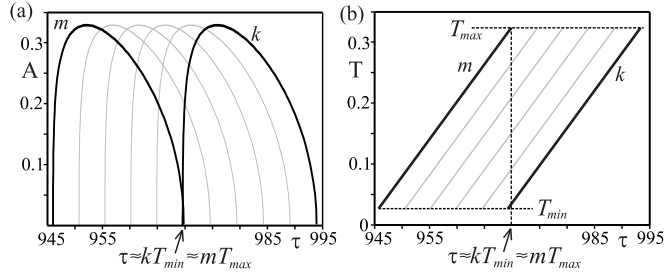}

\caption{Illustration to the derivation procedure of the formula for the number
of overlapping branches at time delay $\tau$. Shown is the delayed
Duffing oscillator with parameter values $a=1.38$,\textbf{ $b=0.4$,
$d=0.3$}, see also Fig.~\ref{fig:Branches-of-periodic}(c). The branches $k$ and $m$
are shown in black. More
details are given in text. \label{fig:Illustration-to-the}}

\end{figure}

Let $m$ be the least number of the branch, which exists at time delay
$\tau$ and $k$ be the largest number, i.e. \[
\tau_{\max}(m,l)\approx\tau\quad\mbox{and}\quad\tau_{\min}(k,l)\approx\tau\]
up to terms of order one, see Fig.~\ref{fig:Illustration-to-the}.
Taking into account (\ref{eq:Taumin}) and (\ref{eq:Taumax}), we
obtain \[
mT_{\max}\approx kT_{\min}\approx\tau,\]
where the approximation sign means that the equality is satisfied
up to the order one terms ($k$ and $m$ are large). All branches
with numbers $m<n<k$ exist at time delay $\tau$, see Fig.~\ref{fig:Illustration-to-the}.
Hence, the number of overlapping branches can be estimated as follows
\begin{equation}
N\approx k-m=k-k\frac{T_{\min}}{T_{\max}}=k\frac{T_{\max}-T_{\min}}{T_{\max}}=\kappa_{1}\tau,\label{eq:lineargrowth}\end{equation}
where the coefficient for this growth is given by \begin{equation}
\kappa_{1}=\frac{T_{\max}-T_{\min}}{T_{\max}T_{\min}}=\frac{1}{T_{\min}}-\frac{1}{T_{\max}}.\label{eq:coeff}\end{equation}
Expression (\ref{eq:lineargrowth}) gives also a lower estimation
for the number of coexisting periodic solutions of system (\ref{eq:dde}).
Indeed, if the branches are folded like in Fig.~\ref{fig:Branches-of-periodic}(a)
then one branch may lead to more than one periodic solution for some
delay values.

In a similar way, one can show that the number of coexisting branches
in CASE 2 can be estimated as \begin{equation}
N\approx\kappa_{2}\tau=\frac{1}{T_{\min}}\tau.\label{eq:growthcase2}\end{equation}
Finally note that there may exist few primary branches, which cannot
be mapped one onto another by the transformation (\ref{eq:nvia0branch}).
In this case, each branch will reappear with the increasing delay.
The growth rate $\kappa$ fin this case is given as a superposition
of the corresponding rates.

\subsubsection*{Turning points on branches. }

The branches may have turning points, which correspond to a fold bifurcation
for the family of periodic solutions. The condition for the branch
$n$ to have a turning point can be written as follows \[
\frac{\partial\tau(n,l)}{\partial l}=0\]
 or, taking into account (\ref{eq:Tmap}) \begin{equation}
1+nT'(l)=0.\label{eq:eq1}\end{equation}
Equation (\ref{eq:eq1}) can be rewritten as \begin{equation}
T'(l)=-\frac{1}{n}.\label{eq:TurningCondition}\end{equation}
With the incresing of branch number $n$, the fold point tends to
some asymptotic value, which is independent on $n$ and given by the
condition $T'(l)=0$.

\subsection{Example: Stuart-Landau oscillator with delay \label{sub:Example:-Stuart-Landau-oscillator}}

In this paragraph we consider the following system \begin{equation}
z'(t)=(\alpha+i\beta)z(t)-z(t)|z(t)|^{2}+z(t-\tau)\label{eq:SL}\end{equation}
with the instantaneous part as the normal form for an oscillator close
to supercritical Andronov-Hopf bifurcation. Such system is called
sometimes Stuart-Landau oscillator \citep{Kuramoto1984,Wolfrum2006}.
The additional term $z(t-\tau)$ accounts for a delayed feedback.
Here $z(t)$ is a complex variable, i.e. the system has essentially
two components, which can be chosen as real and imaginary parts of
$z(t)$. 

Due to symmetry properties of this system, some of its periodic solutions
can be found analytically in the form of rotating waves $Ae^{i\omega t}$.
Substituting this rotating wave into (\ref{eq:SL}), we obtain the
equation \[
i\omega=(\alpha+i\beta)-A^{2}+e^{-i\omega\tau},\]
 which leads to the following expressions for the amplitude $A$ and
frequency $\omega$\begin{equation}
A=\sqrt{\alpha+\cos\omega\tau}.\label{eq:SL-Aomega}\end{equation}
\begin{equation}
\omega=\beta-\sin\omega\tau,\label{eq:SL-omega}\end{equation}
For the purposes of this paper, let us rewrite (\ref{eq:SL-Aomega})
and (\ref{eq:SL-omega}) in terms of the amplitude $A$ and the period
$T=2\pi/\omega$:\begin{equation}
A=\sqrt{\alpha+\cos\left(2\pi\frac{\tau}{T}\right)}.\label{eq:SL-AT}\end{equation}
\begin{equation}
T=\frac{2\pi}{\beta-\sin\left(2\pi\frac{\tau}{T}\right)},\label{eq:SL-T}\end{equation}
The expressions (\ref{eq:SL-AT}) and (\ref{eq:SL-T}) are invariant
under the change $\tau\to\tau+nT$, which reflects the fact that solutions
on different branches are identical. 

\begin{figure}
\includegraphics[width=8cm]{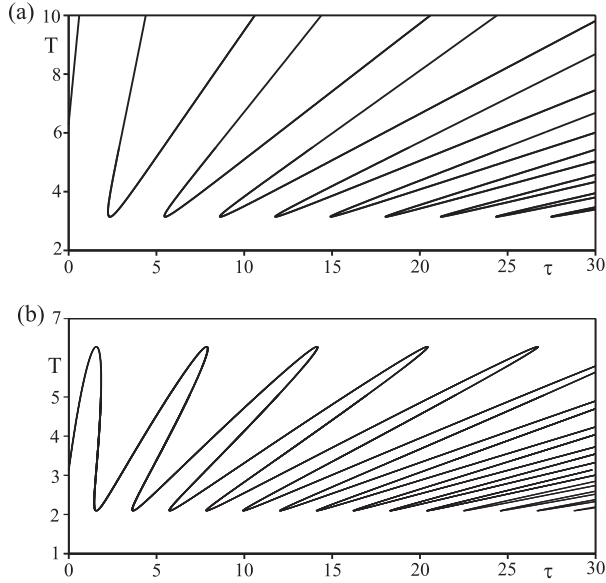}

\caption{Branches of periodic solutions for the Stuart-Landau oscillator with
delay, which are given analytically in (\ref{eq:SLbranches1}) --
(\ref{eq:SLbranches2}); (a) $\beta=1$; (b) $\beta=2$. \label{fig:SL}}

\end{figure}

Although the period $T(\tau)$ along the branch is given in the implicit
way by (\ref{eq:SL-T}), one can obtain an explicit parametric representation
of the branches with respect to $T$ and $\tau$. For this, we introduce
an additional parameter $\psi=2\pi\tau/T$. With the help of this
parameter, solutions of (\ref{eq:SL-T}) can be written as follows
\begin{equation}
T(\psi)=\frac{2\pi}{\beta-\sin\psi},\label{eq:SLbranches1}\end{equation}
\begin{equation}
\tau(\psi)=\frac{\psi+2\pi n}{\beta-\sin\psi}.\label{eq:SLbranches2}\end{equation}
Now the branches can be easily plotted by varying parameter $\psi$.
For $\beta=1$ the branches are unbounded, see Fig.~\ref{fig:SL}(a),
and for $\beta=2$ they are bounded, see Fig.~\ref{fig:SL}(b). Moreover,
for $\beta=2$ neighboring branches are connected.

As it is expected, the coexistence of multiple periodic solutions
grows with the increasing of delay. The number of coexisting branches
can be estimated using (\ref{eq:lineargrowth}) and (\ref{eq:growthcase2})
as \[
N\approx\frac{1}{T_{\min}}\tau=\frac{\beta+1}{2\pi}\tau\]
in the case $0<\beta<1$ and \[
N\approx\left(\frac{1}{T_{\min}}-\frac{1}{T_{\max}}\right)\tau=\frac{\tau}{\pi}\]
for $\beta>1$. Taking into account folding of the branches, the number
of periodic solutions grows twice as fast with the increasing of the
delay, i.e. with the rates $(\beta+1)\tau/\pi$ and $2\tau/\pi$ respectively.

\section{Stability of the reappearing periodic solutions, long delay issues
\label{sec:Stability-of-the}}

\subsection{Some elements of the stability theory for periodic solutions}

Let us introduce necessary notations and shortly remind basic elements
of the stability theory \citep{Hale1977} for periodic solutions of
(\ref{eq:dde}). 

The linearization of (\ref{eq:dde}) around some periodic solution
$p(t)$ with a period $T$ has the following form\begin{equation}
\xi'(t)=A(t)\xi(t)+B(t)\xi(t-\tau),\label{eq:linearization}\end{equation}
where $A(t)=D_{1}f(p(t),p(t-\tau))$ and $B(t)=D_{2}f(p(t),p(t-\tau))$
are $T$- periodic matrices $n\times n$. Here $D_{1}$ and $D_{2}$
denote partial derivatives with respect to the first and the second
argument respectively. Any solution of (\ref{eq:linearization}) with
some initial function $\varphi(t)$ can be represented as $x(t;s,\varphi)=\Psi(t;s)\varphi$,
where $\Psi(t;s)$ is the evolution operator \citep{Hale1977}. The
monodromy operator is introduced as the evolution operator evaluated
at the period \[
U=\Psi(T;0).\]

Stability of the periodic solution $p(t)$ is determined by a countable
set of \emph{characteristic multipliers} \citep{Hale1977,Hale1993}
$\mu_{j}$, $j=1,2,\dots$, which are defined by the spectrum of $U$.
The corresponding characteristic exponents are given as $\lambda_{j}=\frac{1}{T}\mathrm{Ln}\mu_{j}$.
A periodic solution is asymptotically stable if all its multipliers
have modulus less than one, or, equivalently, real parts of all characteristic
exponents are negative. A bifurcation occurs when a multiplier crosses
unitary circle as a parameter change. 

Practically, characteristic multipliers and stability of a periodic
solutions can be computed using DDE-Biftool software.

\subsection{Stability of periodic solutions versus delay }

\begin{figure}
\includegraphics[width=8cm]{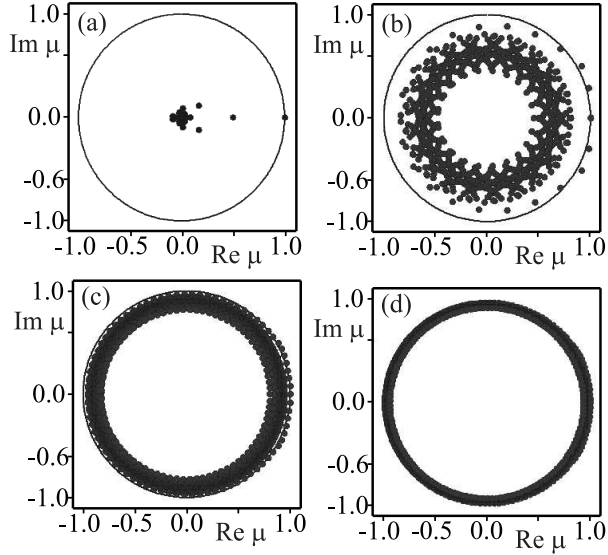}

\caption{Largest characteristic multipliers of a periodic solution of delayed
Duffing oscillator for different values of delay. Different figures
correspond to the same solution on different branches: (a) branch
$n=2$ ($\tau=2.2$); (b) $n=20$ ($\tau=23.7$); (c) $n=80$ ($\tau=91.6$),
(d) $n=140$ ($\tau=159.5$). Other parameters are $a=0.5$, $b=0.6$,
$d=0.06$. \label{fig:Characteristic-multipliers-numerics1}}

\end{figure}

Considering $\tau$ as the control parameter, stability properties
of periodic solutions change as $\tau$ is varied. In general, it
is a challenging problem to find their stability, especially for larger
$\tau$. Figure~\ref{fig:Characteristic-multipliers-numerics1} shows
largest characteristic multipliers (with the largest modulus) for
a periodic solution of delayed Duffing oscillator (\ref{eq:duffing})
for different delay times. Even though the solution is the same, its
stability properties change as one moves from one branch to another.
One can clearly observe that more and more multipliers come close
to the unitary circle making the problem more and more degenerate
and numerically stiff for larger delay. 

In the following we propose an analytical technique, which overcomes
the appeared difficulty and shows how to approximate the characteristic
multipliers for larger delay values. In particular we show that the
characteristic multipliers have similar properties to that of the
eigenvalues of stationary states for systems with large delay \citep{Yanchuk2005a,Wolfrum2006,Yanchuk2006a}. 

The linearization of equation (\ref{eq:dde}) around the solution
$x_{n}(t;\tau)$ on $n$-th branch (this solution coincides with $x_{0}(t;l)$)
at time-delay $\tau=l+nT(l)$ has the form \begin{equation}
\xi'(t)=A(t;l)\xi(t)+B(t;l)\xi(t-\tau),\label{eq:linearl-tau}\end{equation}
where \begin{eqnarray}
A(t;l) & = & D_{1}f(x_{0}(t;l),x_{0}(t-l;l)),\label{eq:AB}\\
B(t;l) & = & D_{2}f(x_{0}(t;l),x_{0}(t-l;l))\nonumber \end{eqnarray}
are $T$-periodic matrices, which depend only on function $f$ and
a shape of the solution $x_{0}(t;l)$. It is important that $A$ and
$B$ do not depend on the branch number $n$ and the time delay $\tau$,
at which the system is considered. This allows us to study stability
properties of periodic solutions asymptotically with $\tau\to\infty$.
Namely, by increasing $\tau$ (or, equivalently, branch number $n$)
we are actually jumping from one branch to another by keeping the
relative position $l$ within the branch fixed, see Fig.~\ref{fig:Mapping-of-periodic}.
As a result, we consider the same periodic solution $x_{0}(t;l)$,
which exists for different infinitely increasing delay values and
we study stability properties of this solution as $\tau$ increases.
In the following, we consider $\tau$ to be continuous parameter and
then apply the obtained results to the countable set $\tau(n,l)=l+nT(l),$
$n=0,1,2,\dots$ of delay values.

\subsection{Pseudo-continuous spectrum}

Here we show that periodic solutions possess a family of characteristic
multipliers, which have the following asymptotic representation \begin{equation}
|\mu_{\omega}|=1+\frac{1}{\tau}\gamma(\omega)+\mathcal{O}\left(\frac{1}{\tau^{2}}\right)\label{eq:PCSperiodic}\end{equation}
with the increasing of delay. Here $\omega$ is a parameter along
the family. Using the analogy to the spectrum of eigenvalues for stationary
solutions \citep{Wolfrum2006,Yanchuk2005a,Yanchuk2006a}, we will
call such spectrum \emph{pseudo-continuous}. Its main features are:
\begin{enumerate}
\item Pseudo-continuous spectrum tends to the critical value $|\mu|=1$
as $\tau\to\infty$;
\item Its stability is determined by the sign of the function $\gamma(\omega)$;
\item For any finite $\tau$, the parameter $\omega$ admits a discrete
countable number of values. As $\tau\to\infty,$ the spectrum tends
to a continuous in the sense that the discrete parameter $\omega$
covers densely the whole interval $[0,2\pi]$. 
\end{enumerate}
In the case, if a periodic solution has only pseudo-continuous spectrum,
its stability for large enough $\tau$ will be uniquely defined by
the function $\gamma(\omega)$. 

\begin{figure}
\includegraphics[width=8cm]{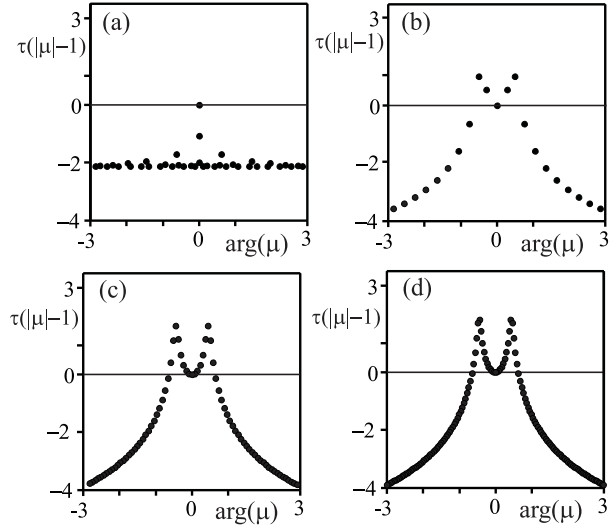}

\caption{Largest characteristic multipliers for a periodic solution of Duffing
oscillator with delay, which are rescaled accordingly to the rule
$\gamma_{j}=\tau(|\mu_{j}|-1)$ (vertical axis) and $\omega=\arg(\mu_{j})$
(horizontal axis). With the increasing of $\tau$, such rescaled spectrum
tends the continuous curve $\gamma(\omega)$ representing the pseudo-continuous
spectrum as described by (\ref{eq:PCSperiodic}). Parameter values
are as in Fig.~\ref{fig:Characteristic-multipliers-numerics1}. \label{fig:Pseudo-continuous-spectrum}}

\end{figure}

Before we give a rigorous proof for the existence of the pseudo-continuous
spectrum, let us illustrate it using our numerical example of Duffing
oscillator (\ref{eq:duffing}). Figure~\ref{fig:Pseudo-continuous-spectrum}
shows approximations for the function $\gamma(\omega)$ by plotting
$\gamma_{j}=\tau(|\mu_{j}|-1)$ versus $\omega=\arg(\mu_{j})$, where
$\mu_{j}$ are numerically obtained largest characteristic multipliers.
One can see that with increasing delay $\tau$, the plot tends to
some continuous curve, which determines stability of the periodic
solution (it is unstable here). Below we give a proof for the existence
of pseudo-continuous spectrum. Those readers, who are not interesting
in details, may skip the following section.

\emph{Proof of the existence of pseudo-continuous spectrum.}\textbf{\emph{ }}

Here we use the theory for linear delay differential equations with
periodic coefficients \citep{Hale1977}, which is analogous to the
Floquet theory for ordinary differential equations. This theory implies
that $\mu=e^{\lambda T}$ is the characteristic multiplier of (\ref{eq:linearl-tau})
if and only if there is a nonzero solution of equation (\ref{eq:linearl-tau})
of the form \begin{equation}
\xi(t)=p(t)e^{\lambda t},\label{eq:Floquetchange}\end{equation}
where $p(t)=p(t+T)$ is periodic. Substituting (\ref{eq:Floquetchange})
into (\ref{eq:linearl-tau}), we obtain \begin{equation}
p'(t)=\left(A(t;l)-\lambda\mathrm{Id}\right)p(t)+e^{-\lambda\tau}B(t;l)p(t-\tau).\label{eq:p1}\end{equation}
Since $p(t)$ is $T$-periodic, we have $p(t-\tau)=p(t-l-nT(l))=p(t-l)$
and system (\ref{eq:p1}) reduces to \begin{equation}
p'(t)=\left(A(t;l)-\lambda\mathrm{Id}\right)p(t)+e^{-\lambda\tau}B(t;l)p(t-l).\label{eq:p2}\end{equation}
where the large parameter $\tau$ occurs only as a parameter in $e^{-\lambda\tau}$.
Thus, the corresponding monodromy operator of (\ref{eq:p2}) \[
U=U(\lambda,e^{-\lambda\tau})\]
depends only on $\lambda$ and $e^{-\lambda\tau}$ smoothly. Since
the linear system (\ref{eq:p2}) possesses the periodic solution $p(t)$
by construction, one of its characteristic multipliers equals to one.
This leads to the following condition on the monodromy operator\[
\left[U(\lambda,e^{-\lambda\tau})-\mathrm{Id}\right]p=0,\]
which must hold for some periodic function $p(t)$. In general case,
this is a co-dimension one condition, i.e. it leads to some characteristic
equation \begin{equation}
F(\lambda,e^{-\lambda\tau})=0\label{eq:generalcondition}\end{equation}
for determining the characteristic exponents $\lambda$. This characteristic
equation allows proving the existence of pseudo-continuous spectrum.
Indeed, substituting \begin{equation}
\lambda=\frac{\gamma}{\tau T}+i\frac{\omega}{T}\label{eq:pseudocont-exp}\end{equation}
into (\ref{eq:generalcondition}), we obtain up to the leading order
\begin{equation}
F(i\frac{\omega}{T},e^{-\frac{\gamma}{T}}e^{-i\frac{\omega}{T}\tau})=0.\label{eq:U2}\end{equation}
New unknown real variables now are $\omega$ and $\gamma$ in Eq.
(\ref{eq:U2}) instead of complex $\lambda$ in Eq. (\ref{eq:generalcondition}).
Further we will proceed similarly to the case of the pseudo-continuous
spectrum for stationary states \citep{Yanchuk2005a}. Let us introduce
the new artificial parameter $\varphi$ instead of rapidly growing
phase $\frac{\omega}{T}\tau$ \begin{equation}
F(i\frac{\omega}{T},e^{-\frac{\gamma}{T}}e^{-i\varphi})=0.\label{eq:U3}\end{equation}
 The newly obtained extended equation (\ref{eq:U3}) can be generically
solved with respect to $\gamma(\omega)$ and $\varphi(\omega)$, since
the equation is complex, i.e. it contains two real equations for two
variables $\gamma$ and $\varphi$. The obtained function $\gamma(\omega)$
is the resulting asymptotic function, which determines the pseudo-continuous
spectrum, see Fig.~\ref{fig:Pseudo-continuous-spectrum}. Coming
back from the extended equation (\ref{eq:U3}) to the original one
(\ref{eq:U2}), we additionally have to take into account the condition
\[
\frac{\omega}{T}\tau=\varphi(\omega)+2\pi k,\quad k=0,\pm1,\pm2,\dots\]
or, equivalently \begin{equation}
\frac{\omega}{T}=\frac{1}{\tau}\varphi(\omega)+\frac{2\pi k}{\tau}.\label{eq:omega}\end{equation}
The last equation (\ref{eq:omega}) determines the discrete values
of $\omega=\omega_{k}$, which correspond to the discrete values of
the pseudo-continuous spectrum $\gamma(\omega_{k})$. As $\tau$ increases,
the set of $\omega_{k}$ covers densely the whole domain of $\omega$. 

As a result, the pseudo-continuous spectrum approaches the continuous
one as $\tau\to\infty$. Note, that the asymptotically continuous
spectrum $\gamma(\omega)$ is determined by a regular system of equations
(\ref{eq:U3}), which no longer contains the large parameter $\tau$.
Finally we remark that characteristic exponents (\ref{eq:pseudocont-exp})
correspond to characteristic multipliers (\ref{eq:PCSperiodic}).

\subsection{Strongly unstable spectrum}

For completeness we show that another type of characteristic multipliers
may appear which have different asymptotics for large $\tau$. These
multipliers are not approaching the threshold value $|\mu|=1$ as
$\tau\to\infty$, but tending to some unstable value \[
\mu\to\bar{\mu},\quad|\bar{\mu}|>1.\]
Since $|\bar{\mu}|>1$, the corresponding periodic solution is unstable.
In the case, if some periodic solution has a strongly unstable multiplier,
close-by solutions will diverge on a time interval much shorter than
the delay $\tau$ simply because the divergence rate is independent
on the delay. 

Strongly unstable spectrum may consist of finite number of multipliers
(less than $n$) and it occurs if and only if the instantaneous part
of (\ref{eq:linearl-tau}), i.e. the system of ordinary differential
equations (ODE) \begin{equation}
\xi'(t)=A(t;l)\xi(t).\label{eq:ode}\end{equation}
is unstable. The unstable Floquet multipliers of this system will
serve as asymptotic values $\bar{\mu}$ for the strongly unstable
characteristic multipliers of the system with delay (\ref{eq:linearl-tau}). 

\begin{figure}
\includegraphics[width=8cm]{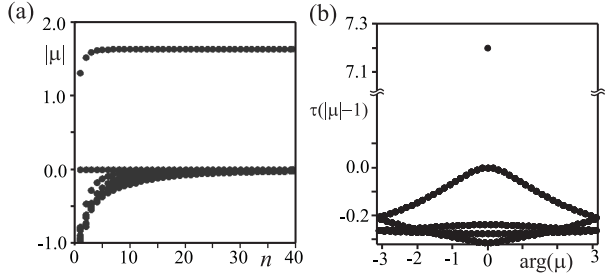}

\caption{Example of a strongly unstable spectrum for a periodic solution of
Duffing oscillator with delay. Single multiplier with large modulus
is the strongly unstable one. The others are belonging to the pseudo-continuous
spectrum. (a) Modulus of largest characteristic multipliers versus
branch numbers (delay is increasing). (b) Rescaled spectrum, i.e.
$\tau\left(|\mu|-1\right)$ versus argument of $\mu$ for largest
characteristic multipliers of a periodic solution on the branch 40.
Note the scale on the vertical axis. Parameter values: $a=0.5$, $b=0.6$,
$d=0.06$. \label{fig:Example-of-strongly}}

\end{figure}

An example of strongly unstable spectrum is shown in Fig.~\ref{fig:Example-of-strongly}.

\emph{Why does strongly unstable spectrum exist?}

The idea of the proof is the following. Assume that there is a characteristic
multiplier $\mu$ with $|\mu(\tau)|>1$, which persists for all $\tau\to\infty$
and do not scale with $\tau$. Then $\mu^{-\tau/T}\to0$ as $\tau\to\infty$
and the delayed term in (\ref{eq:p2}) \begin{equation}
e^{-\lambda\tau}B(t;l)p(t-l)=\mu^{-\tau/T}B(t;l)p(t-l)\label{eq:delayedterm}\end{equation}
becomes exponentially small comparing with the term $\left(A(t;l)-\lambda\mathrm{Id}\right)p(t)$.
In fact, choosing large enough $\tau$, it can be made arbitrary small.
Therefore, the equation (\ref{eq:p2}) can be formally reduced to
the ODE \begin{equation}
p'(t)=\left(A(t;l)-\lambda\mathrm{Id}\right)p(t).\label{eq:pode}\end{equation}
The condition for (\ref{eq:pode}) to have a multiplier equal to one
reduces to the equation \begin{equation}
\det\left[U_{0}-e^{\lambda T}\mathrm{Id}\right]=0,\label{eq:solsol}\end{equation}
where $U_{0}$ is the monodromy matrix of the ODE (\ref{eq:ode}).
Since the solutions of (\ref{eq:solsol}) are multipliers of (\ref{eq:ode}),
strongly unstable multiplier will approach an unstable multiplier
$\bar{\mu}=e^{\lambda T}$ of (\ref{eq:ode}).

\section{Asymptotic stability of branches \label{sec:Asymptotic-stability-of}}

\begin{figure}
\includegraphics{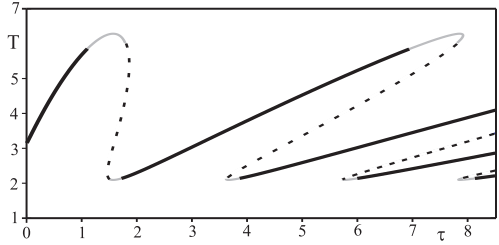}

\caption{This figure illustrates how branches of periodic solutions have asymptotically
well defined stability structure. Black solid lines correspond to
asymptotically stable parts of the branch for large enough delay ($-1.22<\psi<1.18$),
dashed are strongly unstable ($1.868<\psi<4.415$) gray parts are
weakly unstable. Parameter values are $\alpha=\beta=2$. \label{fig:This-figure-illustrates}}

\end{figure}

Let us discuss the main consequences, which follow from the existence
of the pseudo-continuous spectrum. First of all, let us note, that
the sequence of periodic solutions $x_{n}(t,\tau)=x_{0}(t,l)$, which
repeat themselves at time delays $l+nT(l)$, has a well defined stability
limit as $n$ increases. This means, that all solutions from this
sequence with large enough $n$ will be stable if the corresponding
pseudo-continuous spectrum is stable and no strongly unstable spectrum
is present. Otherwise, the corresponding solutions will be unstable.
In other words, in the limit of large delay, the branches of periodic
solutions have well defined stability structure, i.e. there will be
some stable part as well as unstable part. The unstable part can be
again splitted into strongly unstable (if there are strongly unstable
multipliers) or weakly unstable (when only pseudo-continuous spectrum
is unstable). The corresponding parts can be described by the parameter
$l$ on the branches. Figure~\ref{fig:This-figure-illustrates} illustrates
this using the Stuart Landau model with $\alpha=2$ and $\beta=2$,
see caption to the figure. 

Taking into account that almost all parts of the branches are eventually
stretching with the increasing of delay, an increasing coexistence
of stable as well as unstable periodic solutions is generally expected
in systems with large delay. The relative fraction of stable solutions
depends on a specific system, more exactly, on the asymptotic spectrum
distribution along the branch.

\section{Computation of the pseudo-continuous spectrum \label{sec:Computation-of-pseudo-continuous}}

The main equation for finding branches $\gamma(\omega)$, to which
the pseudo-continuous spectrum tends to, is given by (\ref{eq:U3}).
As follows from the previous section, the equivalent problem can be
formulated as follows:

For any given $\omega$, find a value of $\gamma=\gamma(\omega)$
and $\varphi(\omega)$ such that the following linear system with
delay \begin{equation}
p'(t)=\left(A(t;l)-i\frac{\omega}{T}\mathrm{Id}\right)p(t)+e^{-\frac{\gamma}{T}-i\varphi}B(t;l)p(t-l)\label{eq:ppp}\end{equation}
has a multiplier 1. Here $l=\tau\,\mathrm{mod}\, T$ and $A$ and
$B$ are determined by linearizing (\ref{eq:dde}) around the given
periodic solution with period $T$; see (\ref{eq:AB}). Equivalently,
the following extended system can be considered

\begin{eqnarray}
x'(t) & = & f(x(t),x(t-\tau)),\nonumber \\
p'(t) & = & \left(D_{1}f(x(t),x(t-l))-i\frac{\omega}{T}\mathrm{Id}\right)p(t)\label{eq:pathfollow1}\\
 &  & +e^{-\frac{\gamma}{T}-i\varphi}D_{2}f(x(t),x(t-l))p(t-l)\nonumber \end{eqnarray}
with the following additional conditions \begin{eqnarray}
p(t) & = & p(t+T),\nonumber \\
x(t) & = & x(t+T),\label{eq:pathfollow2}\\
\|p(t)\| & = & 1,\nonumber \end{eqnarray}
where the first two conditions ensure periodicity and the second one
ensures that $p(t)$ is nontrivial. This is a typical non-stiff continuation
problem, which no longer includes the large parameter $\tau$. Standard
continuation algorithms should be used in order to find the functions
$\gamma(\omega)$ and $\varphi(\omega)$. 

The implementation of the continuation algorithm will be discussed
elsewhere \citep{Sieber}. Instead, we discuss here cases, at which
the above problem can be significantly simplified. 

Case 1: $\tau\,\mathrm{mod}\, T=0$. This situation appears if system
(\ref{eq:dde}) has a periodic solution at $\tau=0$. In this case,
the equation (\ref{eq:ppp}) is reduced to the ODE \begin{equation}
p'(t)=\left(A(t;l)-i\frac{\omega}{T}\mathrm{Id}+e^{-\frac{\gamma}{T}-i\varphi}B(t;l)\right)p(t)\label{eq:odeode}\end{equation}
and the equivalent problem for finding functions $\gamma(\omega)$
and $\varphi(\omega)$ reduces to the finite-dimensional ODE continuation
problem \citep{Dhooge2006,Doedel2006}.

\begin{figure}
\includegraphics[width=8cm]{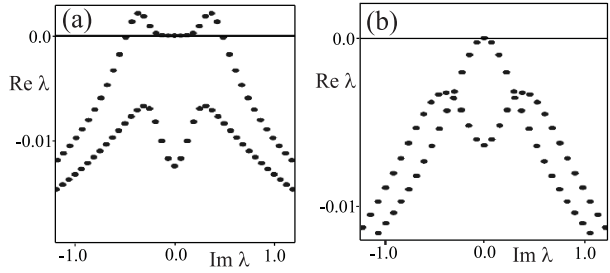}

\caption{Illustration of pseudo-continuous spectrum for external cavity modes
of the Lang-Kobayashi system (adapted from \citep{Yanchuk2004b}).
\label{fig:Illustration-of-pseudo-continuous}}

\end{figure}

Case 2: system (\ref{eq:dde}) has an additional phase shift symmetry.
Examples of such systems are Stuart-Landau oscillator (\ref{eq:SL})
or Lang-Kobayashi system \citep{Lang1980,Alsing1996,Heil1998,Haegeman2002,Wolfrum2006}
Spectrum of some external cavity mode for the Lang-Kabayashi system
is shown in Fig.~\ref{fig:Illustration-of-pseudo-continuous}. Other
examples can be found in \citep{Yanchuk2006a}.

\section{Conclusions}

To conclude, we have investigated properties of periodic solutions
of systems with delay. In particular, we have shown that 
\begin{itemize}
\item Periodic solutions of systems with delay are organized in infinitely
many branches.
\item The branches of periodic solutions can be obtained as the mapping
(\ref{eq:nvia0branch}-\ref{eq:Tmap}) of a primary branch on an appropriate
interval of $\tau$. From practical points of view, it is sufficient
to calculate only the primary branch.
\item The branches are eventually becoming wider with the increasing of
$\tau$, i.e. they occupy larger $\tau$ interval. As a result, the
multistability of periodic solutions grows as delay increases.
\item This growth of the multistability is linear (\ref{eq:lineargrowth})
and the corresponding estimation is given in (\ref{eq:coeff}) and
(\ref{eq:growthcase2}).
\item One can effectively study asymptotic stability properties of periodic
solutions as $\tau\to\infty$. 
\item As $\tau$ becomes larger, the spectrum of characteristic multipliers
of periodic solutions is splitted into two parts: pseudo-continuous
spectrum and strongly unstable spectrum.
\item The main properties and implications of pseudo-continuous spectrum
are explained. In particular, pseudo-continuous spectrum controls
the destabilization of periodic solutions. It shows also that the
destabilization mechanism of periodic solutions should be similar
to that of spatially extended systems \citep{Wolfrum2006,Giacomelli1996}.
Moreover, it implies that the dimensionality of unstable manifolds
of periodic orbits grows linearly as delay increases.
\item Strongly unstable spectrum is present when the instantaneous part
of the linearization around the periodic solution is unstable. In
this case, the feedback plays minor role, see also \citep{Kinzel}.
\end{itemize}
In our paper we also outlined the algorithm for practical finding
of the pseudo-continuous spectrum, although the numerical implementation
of the corresponding path-following algorithm (see (\ref{eq:pathfollow1})--(\ref{eq:pathfollow2}))
will be discussed elsewhere \citep{Sieber}. 

The authors acknowledge the support of DFG Research Center Matheon
{}``Mathematics for key technologies'', DAAD cooperation project
D0700430 and Department of International Co-operation of Poland project
DWM/N97/DAAD/2008. We also acknowledge valuable discussions with Matthias
Wolfrum, Tomasz Kapitaniak, and Andrzej Stefanski.

\bibliographystyle{apsrev}

%\bibliography{C:/Users/Vista/Documents/Ybibliography/Ybibliography}

\end{document}